\documentstyle[art12,12pt,epsf]{article} 
\def\mathrm{\rm}
\newcommand{\ra}{\mbox{$\rightarrow$}}

\newcommand{\cbar}{\mbox{${\overline c}$}}
\newcommand{\bbar}{\mbox{${\overline b}$}}
\newcommand{\bbbar}{\mbox{$b\bbar$}}
\newcommand{\ccbar}{\mbox{$c\cbar$}}
\def\ccbar{\mbox{${c\cbar}$}}
\newcommand{\qbar}{\mbox{${\overline q}$}}
\newcommand{\qqbar}{\mbox{$q\qbar$}}

\newcommand{\ee}{\mbox{${e^{+}e^-}$}}
\newcommand{\effb}{\mbox{${\epsilon_b}$}}
\newcommand{\effc}{\mbox{${\epsilon_c}$}}
\newcommand{\effu}{\mbox{${\epsilon_{uds}}$}}
\newcommand{\lamb}{\mbox{${\lambda_b}$}}

\newcommand{\Mtag}{\mbox{${\cal M}$}}
\newcommand{\glubb}{\mbox{${g\ra{b\overline{b}}}$}}
\newcommand{\glucc}{\mbox{${g\ra{c\overline{c}}}$}}

\newcommand {\bea} {\begin{eqnarray}}
\newcommand {\eea} {\end{eqnarray}}
\newcommand {\beq} {\begin{equation}}
\newcommand {\eeq} {\end{equation}}

\newcommand{\RBFGAM}{0.2142}
\newcommand{\DRBFST}{0.0034}
\newcommand{\DRBFSYS}{0.0015}
\newcommand{\DRBFRC}{0.0002}

\newcommand{\EFRB}{35.3\%}
\newcommand{\DEFRB}{0.6\%}
\newcommand{\MCLAMB}{0.59\%}
\newcommand{\Nbbesel}{72074}
\newcommand{\PRBMR}{98\%}
\newcommand{\EFRBMR}{28\%}

\begin{document}   


\begin{titlepage}
  \begin{flushright}
  SLAC-PUB-7481\\
  August 12, 1997
  \end{flushright}
\begin{center}
 {\Large{\bf  A Measurement of 
$R_b$ using a Vertex Mass Tag}}
\end{center}

\begin{center}
{\bf The SLD Collaboration*}
 
{\it Stanford Linear Accelerator Center}
 
Stanford University, Stanford, CA 94309

Submitted to Physical Review Letters

\end{center}
\bigskip
\vspace{2.cm} 
\begin{abstract}
\noindent

We report a new measurement of
$R_b=\Gamma_{Z^o\rightarrow b\overline{b}}/
\Gamma_{Z^0\rightarrow hadrons}$ using a double tag technique, where  
the $b$ hemisphere selection is based on the reconstructed mass of the
$B$ hadron decay vertex.
The measurement was performed using a sample of 
130k hadronic $Z^0$ events, collected with the
SLD at the SLC.
The method utilizes the 3-D vertexing
abilities of the  CCD pixel vertex detector and
the small stable SLC beams to obtain a high $b$-tagging
efficiency and purity.
We obtain 
$R_b=\RBFGAM\pm\DRBFST(stat.)\pm\DRBFSYS(syst.)\pm\DRBFRC(R_c)$.
\end{abstract}
\vfill
\end{titlepage}
\clearpage

We report a new measurement of $R_b$, the fraction of 
$Z^0\rightarrow b\overline b$ events in hadronic 
$Z^0$ decays, collected with  the SLD at
SLC using a  mass tag technique.
The ratio $R_b$ is of special interest as a test of the Standard Model
(SM), since it is sensitive to  possible new physics effects which modify
the  radiative corrections to $Zb\overline{b}$ vertex. 
The vertex corrections are isolated because $R_b$
is a ratio between two hadronic rates, hence propagator
(oblique), radiative and QCD corrections common to all quark 
flavors mostly cancel. 
Recent measurements yielded a  world average $R_b$ value 3$\sigma$ higher 
than that predicted by the SM~\cite{MIKE}. 
Previous measurements~\cite{RB-lifetime} selected $\bbbar$ events based upon
mainly the long $B$ hadron
lifetime and were limited systematically  by contamination in the
sample from residual $\ccbar$ events.
To avoid this limitation our $b$-tag exploits
the large $B$ mass,
since the mass distribution has a very small charm contamination
beyond the charm mass cutoff.
Taking advantage of  SLD's precise 3-D vertexing capability
and the small and stable SLC beam spot, 
we achieve a very efficient and pure $b$ selection.
We use a  self-calibrating double tag technique~\cite{RB-lifetime}, 
which allows one to
measure both $R_b$ and the $b$-tag efficiency, $\effb$,
simultaneously.

This measurement  is performed using approximately 130K 
$\ee \to Z^0\to\qqbar$ 
events collected during 1993-95. 
A detailed description of the detector can be found
 elsewhere~\cite{rbprd}. 
We used the information from charged particle tracks measured
with the CCD pixel Vertex Detector (VXD) along with the
Central Drift Chamber.
The event selection and the determination of the thrust axis
use the  the energy deposits measured with
the Liquid Argon Calorimeter.

The luminous region of the SLC interaction point ($IP$)
has a size of  about ($1.5 \times 0.8$)~$\mu$m in the x,y plane 
transverse to the beam direction and  700~$\mu$m along the beam direction. 
We  use  the average
$IP$ position 
of small groups of sequential hadronic events to determine the 
 primary vertex ($PV$)  in the x-y 
plane.
The longitudinal position of the $PV$ is determined
for each event individually~\cite{rbprd}.
This results in a $PV$ position measurement with uncertainties of 7~$\mu$m 
transverse to the beam axis and 35~$\mu$m (52~$\mu$m for $\bbbar$
events) along the axis. 
The measured track impact parameter resolution 
is $\sigma_{r\phi}[\mu m]=11\oplus 70/p \sin^{3/2} \theta$,   
$\sigma_{rz}[\mu m]=37\oplus 70 /p \sin^{3/2} \theta $ where $p$ is
the track momentum expressed in GeV/c. 

The hadronic event selection is based on charged track multiplicity and 
track visible energy requirements as
described in Ref.~\cite{rbprd}.
The event selection is studied with Monte Carlo (MC) events generated
using JETSET 7.4 event generator~\cite{JETSET}, 
where the $B$ hadron decays are simulated using
a model tuned to current $B$ and $D$ decay data~\cite{CLEO-QQ}.
A plane transverse to the thrust axis is used to divide the event into two
hemispheres.
In order to ensure that the events
are well contained  within
the acceptance of the VXD,
the polar angle of the thrust is required to be  within
$\left| \cos\theta_{thrust}\right| < 0.71$.
In addition, to ensure the event hemisphere division is sensible and to reduce 
the contribution from events containing $g\to\bbbar$,
we require that the event contain no more 
than three jets (defined using charged 
tracks  and the JADE
algorithm~\cite{JADE} 
with $y_{cut}$=0.02).
A total of $\Nbbesel$ events were selected.

In each event well measured tracks~\cite{rbprd} are used to search for
a secondary vertex ($SV$).
The $SV$ are found  by searching 
for areas of high track overlap density
from the individual track resolution functions,   
in 3-D co-ordinate space~\cite{DJNIM}.
The $SV$ are required to be separated from the $PV$ by at least
1~mm and to contain at least two tracks each with a 3-D impact
parameter with respect to the $IP$
$\geq 130\mu$m, ensuring they originate from the decay of a 
particle with relatively 
long lifetime.
Simulation studies show that secondary vertices are found 
in 50\% of all $b$
hemispheres, in 15\% of the charm and $<1\%$ of
the light quark hemispheres~\cite{DJNIM}. 

\begin{figure}[htb]
\epsfysize=13cm.
\epsffile[0 150  312 700]{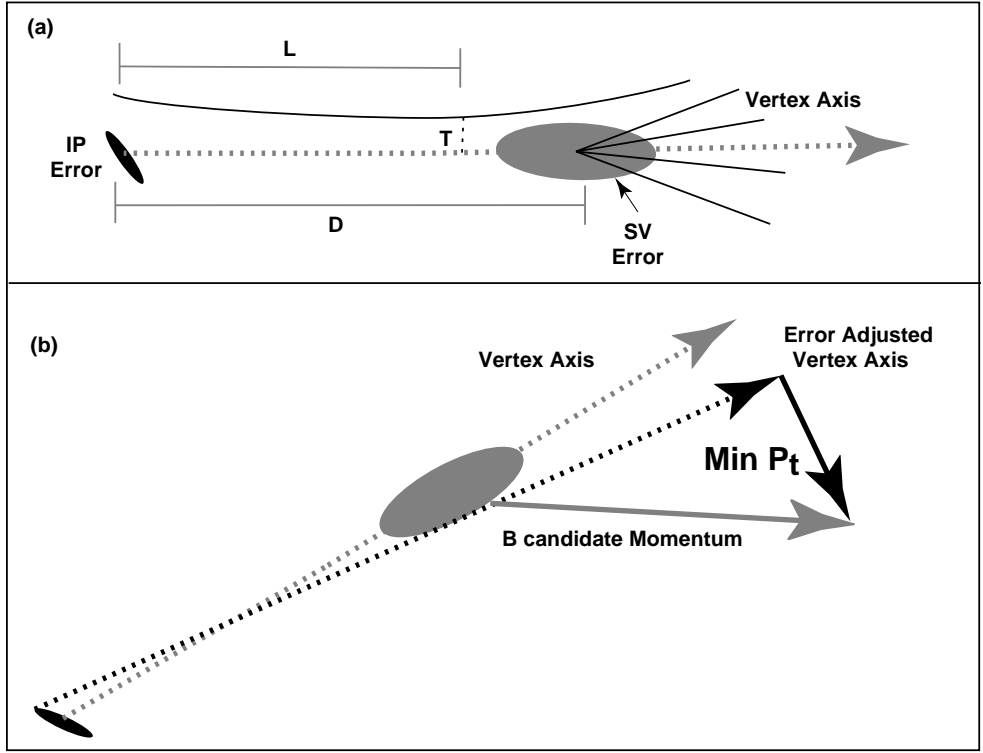}
\caption{(a) An illustration of the SV track attachment criteria. (b)
Illustration of the  $P_t$ derivation.}
\label{fig:illustration}
\end{figure}

\begin{figure}[htb]
\epsfysize=13cm.
\epsffile[0 0  312 500]{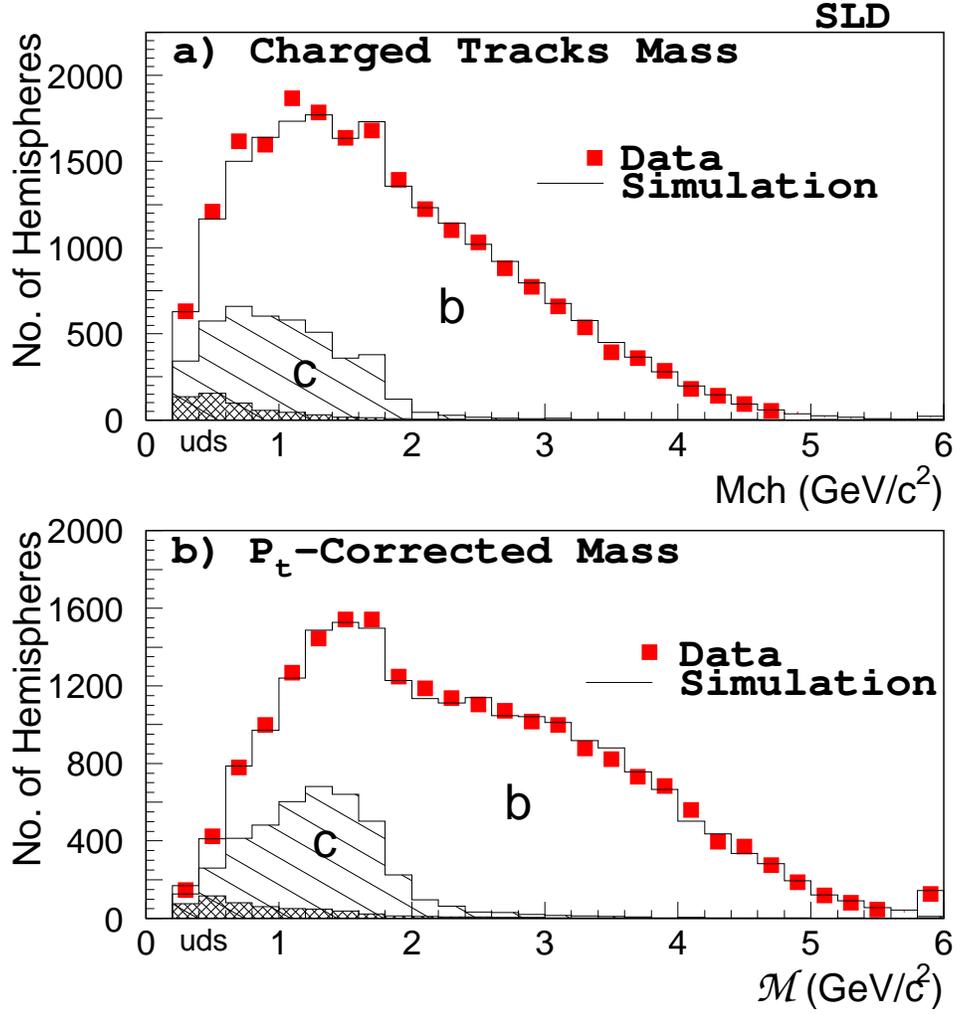}
\caption{Distribution of (a) $M_{ch}$ and (b) $P_t$ corrected
mass, ${\cal M}$,  for data (points) and MC which includes a 
breakdown of the $b$, $c$ and $uds$
contributions (open, hatched and crosshatched histograms respectively).}
\label{fig:mass}
\end{figure}

Due to the cascade structure of the $B$ decay, not all the tracks
in the decay chain will come from a common decay point, thus the $SV$
is incomplete.
We improve our estimate of the 
$B$ decay vertex mass 
by attaching to the $SV$ additional tracks that
are consistent with the hypothesis of originating
from the same $SV$.
We illustrate this in 
Fig.~\ref{fig:illustration}(a). 
We define the
vertex axis to be the   straight line between
the $PV$ and $SV$  centroids.
For each
track not in the $SV$ the 3-D distance of closest approach, $T$, and the distance from the
$PV$ along the vertex axis to this point, $L$, are calculated. 
Tracks with $T<1$~mm and $L/D>0.25$, where
$D$ is the distance from the $PV$ to the $SV$ are attached to the
$SV$ to form 
a $B$ decay candidate.
The invariant mass, $M_{ch}$, of the $B$ candidate is obtained 
assuming each track has the
mass of a charged $\pi$; the distribution of $M_{ch}$ is shown in  
Fig.~\ref{fig:mass}(a).
We require $M_{ch}$ to be well above  the charm mass, $M_{ch}> 2$
~GeV/$c^2$, results in a $b$ hemisphere tagging efficiency of 
$\EFRBMR$ with a purity of $\PRBMR$. 

We improve the $b$ tagging efficiency by applying a kinematic
correction to the calculated $M_{ch}$.
Due to the neglect of information about the neutral particles in the
decay, the $SV$ flight path and the $SV$ momentum vector are typically acollinear.
In order to compensate for the acollinearity 
we correct $M_{ch}$ using
the minimum missing
momentum ($P_t$) transverse to the $SV$ flight
path.
To reject non-$\bbbar$ events with an artificially large $P_t$ 
due to detector resolution effects, 
we define  $P_t$  with respect to  
a vector tangent to the error boundaries of both the $PV$ and the
$SV$, such  that $P_t$ is
minimized (see Fig.~\ref{fig:illustration}(b)). 
The  ability to make this
minimal correction is most effective at  SLD  due to the small and stable
beamspot of the SLC and the high resolution vertexing.
We then define the $P_t$-corrected mass, 
$\Mtag = \sqrt{M_{ch}^{2}+P_{t}^{2}} + |P_{t}|$,
and require 
$\Mtag \leq 2 \times M_{ch}$ to 
reduce the contamination from fake vertices in light quark events.
The distribution of $\Mtag$ is shown in Fig.~\ref{fig:mass}(b).
By requiring   $\Mtag > 2 GeV/c^2$ we
significantly raise our
 $b$-tag efficiency, yielding  $\effb=\EFRB$ 
for the same purity.

\begin{figure}[htb]
\epsfysize=13cm.
\epsffile[70 220  350 550]{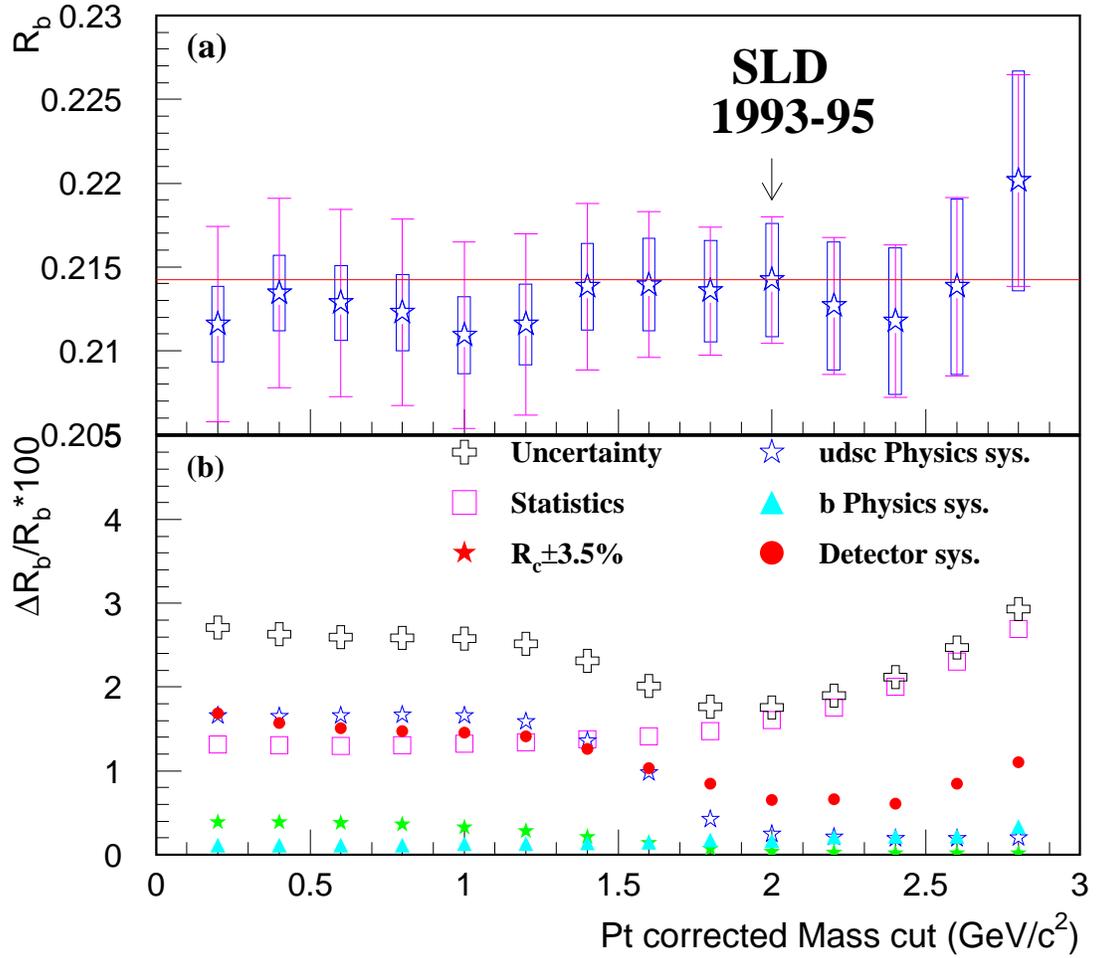}
\caption{(a) Variation of $R_b$ with the ${\cal M}$ cut. 
The statistical error is represented by the boxes and the outer
bars show the sum in quadrature of the statistical and systematic errors.
(b) $R_b$ statistical and systematic uncertainties versus ${\cal M}$ cut.}
\label{r2bsyspic}
\end{figure}

We measure $R_b$ and $\effb$  by counting 
the fraction of the event sample containing one tagged hemispheres,
$F_s$,   
and  the fraction containing both hemispheres tagged, $F_d$:
.
\bea
R_b &= &\frac{(F_s-R_c(\effc-\effu)-\effu)^2}
{F_d-R_c(\effc-\effu)^2+\effu^2-2F_s \effu-\lamb R_b(\effb-\effb^2)}, 
\nonumber \\
\effb &= &\frac{F_d-R_c \effc(\effc-\effu)-F_s \effu-\lamb R_b(\effb-\effb^2)}
{F_s-R_c(\effc-\effu)-\effu}. \nonumber
\eea
The only term dependent upon $B$ production and decay
modeling is the $b$ hemisphere tagging correlation, 
$\lamb=\frac{\effb^{double}-\effb^2}{\effb-\effb^2}=\MCLAMB$, where we
have used the simulation
to estimate $\lamb$.
Estimates of the hemisphere tagging rates of light quarks,
$\effu=0.06\%$  and charm quarks, $\effc=0.69\%$,  are also
derived from the simulation, and we assume 
$R_c=\frac{\Gamma_{Z^0\to c\bar{c}}}
{\Gamma_{Z^0\to q\bar{q}}}=0.171$.
We measure  $R_b=\RBFGAM\pm\DRBFST_{stat.}$
which includes a correction of  +0.0003 for the $\ee\to\gamma\to\bbbar$ contribution
as calculated by ZFITTER~\cite{ZFITTER}.
The measured value of $\effb=\EFRB \pm \DEFRB$ 
is in good agreement with the MC estimate of 35.5\%.
As a cross-check we repeated the measurement using 
different \Mtag\ cuts; the results are summarized in  Fig.~\ref{r2bsyspic}(a).
The $R_b$ results are
consistent for values of $0<\Mtag_{cut}<3$~GeV/$c^2$. 

The systematic uncertainty on $R_b$, given in detail in Table~1, 
results from a combination of detector related 
effects and physics uncertainties in the simulation which 
affect our estimates of $\effc$, $\effu$  and $\lamb$. 
The physics systematic errors are assigned by comparing the
nominal simulation distributions with an alternative set of distributions
which reflect the uncertainties in the world average measurements 
of the MC physics parameters~\cite{EWWG96}.
The two significant sources of systematic errors 
from light quark events come from the uncertainties in
long lived strange particle production and gluon splitting into heavy 
quark pairs.   
The effects of strange particle production  
are studied by varying the $s\overline{s}$ production probability in jet 
fragmentation. 
The \glubb\ and \glucc\ production rates are varied
based upon the OPAL \glucc\ measurement~\cite{OPALglucc}
and the theoretical prediction for the ratio 
$\glubb/\glucc$~\cite{EWWG96}.

The various charmed hadron production rates 
and fragmentation parameters in $Z^0$ decays
are varied within the present LEP measurement 
errors. Charmed hadron fragmentation is studied by varying the 
average scaled energy 
$\langle x_E \rangle$
in the  Peterson fragmentation function~\cite{Peterson},
as well as by studying the 
difference between the Peterson and 
Bowler models~\cite{Bowler} for the same values of
$\langle x_E \rangle$. 
Charmed hadron decay lifetimes are varied according to the world average 
measurement errors~\cite{PDG94}.
The charmed hadron decay charged multiplicity and $K^0$ production rate 
systematic uncertainties are based on measurements by  
Mark-III~\cite{MK3-DDCY}. 
Charmed hadron decays with fewer neutral particles have higher charged mass
and
are therefore more likely to be tagged.
Thus, an additional systematic uncertainty  is estimated by
varying the rates of charmed hadron decays with no $\pi^0$s by $\pm10\%$.

The $B$ decay modeling uncertainty enters via 
the $\lamb$ estimation.
It is studied by varying the $B$ lifetime,
 $B$ baryon production rate, $B$ fragmentation function 
and the $B$ 
decay charged multiplicity
in a similar manner as for the charm systematic studies. 
Simulation uncertainties which affect the tagging efficiency are studied
by comparing the angular distribution of the $b$-tagging rate between
data and simulation
and a systematic error is assigned to the difference.
Hard gluon radiation effects are 
estimated from $\pm 30\%$ variation of the
fraction of simulation events where both $B$ hadrons are contained within 
the  same hemisphere and a hard gluon is in the other.
Another systematic error is assigned to effects of $B$ hadron momentum
correlation between the two hemispheres, mostly due to soft gluon
radiation and fragmentation effects, which in turn translate to a
$b$-tagging efficiency correlation,. This is estimated by comparing
the $B$ momentum correlation in  the HERWIG~\cite{HERWIG} and 
JETSET~\cite{JETSET} event generators.  

As a cross check we  decomposed  the efficiency 
correlation into
an independent set of components 
which represent all sources of correlation between the two $b$ 
hemispheres.
The components we have  studied and their contributions are: the $PV$ measurement
 (-0.02\%), the track resolution effect on the $IP$
determination (+0.04\%), the detector
non-uniformity via the tagging  
angular distribution
dependence (+0.49\%),
the momentum distribution of the $B$ hadron in each hemisphere (+0.08\%) 
and the effect of hard gluon
emission forcing the two $B$ hadrons into one hemisphere (+0.07\%).
The estimated $\lamb$ $(0.59 \pm 0.11)\%$ and that from the sum of
the components (0.67\%)
 are in  good agreement.

A major source of detector systematic uncertainty is due to the discrepancy in
modeling the track impact parameter resolution, mainly along
the beam axis.
In the simulation track $z$ impact parameters are smeared using a
random Gaussian distribution of  width
20~$\mu$m /$\sin\theta$,
as well as adjusted for $z$ impact parameter mean position shifts to
match the data.
The full difference in $R_b$
between the nominal and resolution-corrected samples is
conservatively assigned to be the resolution 
systematic error.
The difference between the measured and simulation charged track multiplicity as a
function of $\cos\theta$ and
momentum is attributed to an unsimulated tracking inefficiency correction.
Both the tracking resolution and efficiency corrections require the
use of a random number generator.
After application of these corrections the result vary slightly with different random
sequences.
These fluctuations are included as an additional MC statistical uncertainty.
The uncertainty on the primary vertex $xy$ location simulation is
estimated from the effect of adding 
a Gaussian tail to the  $ {\rm IP}$ distribution of
 $100\mbox{$\:\mu${\rm m}}$ width for
$0.5\% $  of the  simulated events.
The simulation shows that the $\leq 3$ jets requirement in the event selection 
favors \bbbar\ over other \qqbar\ events which biases our 
measurement by +0.55\%.
We verified this bias in the data, by measuring $R_b$ with and without
applying the  $\leq 3$ jet criterion and found that our measured $R_b$
value only changed by 0.0001, which is consistent with a statistical fluctuation
but nevertheless was included as a systematic error. 
Another bias of $+0.26 \pm 0.12\%$ is introduced by the other event selection
criteria, thus the combined bias is $+0.82\pm0.13\%$ 
and was corrected.
Fig.~\ref{r2bsyspic}(b) shows  the statistical and the  detector, 
physics and $R_c$ systematic uncertainties  
versus the minimum ${\cal M}$ 
cut.

In summary we have  measured:
\bea
R_b=\RBFGAM\pm\DRBFST_{stat.}\pm\DRBFSYS_{syst.}\pm\DRBFRC_{R_c}
\nonumber
\eea
which includes a correction of +0.0003 for the $\ee\to\gamma\to\bbbar$ contribution.
This value supersedes our previous $R_b$ 
measurements~\cite{rbprd} and is in good agreement with the SM prediction of 0.2155.
A new high precision
measurement has recently been reported by ALEPH~\cite{ALEPHNEW}, 
which also incorporates mass
information to improve a lifetime-based probability tag.
With the new SLD and LEP measurements the gap between the  SM
prediction of $R_b$ and the world average has narrowed.    

We thank the personnel of the SLAC accelerator department and the
technical staffs of our collaborating institutions for their
outstanding efforts on our behalf.
This work was supported by the U.S. Department of Energy 
and National Science Foundation, 
the UK Particle Physics and Astronomy Research Council,
the Istituto Nazionale di Fisica Nucleare of Italy,
the Japan-US Cooperative Research Project on High Energy Physics,
and the Korea Science and Engineering Foundation. 

\bibliographystyle{plain}

\begin{thebibliography}{60}
{\leftskip 1.2 cm 
\rightskip 1.2 cm

 \bibitem{MIKE} M.~ Hildreth 
                     proceedings of XXXI Rencontres de Moriond {\sl 
                     ``Electroweak Interactions and Unified Theories''},
J. Tran Than Van ed., (Editions Frontiers 1996) 147.                  

\bibitem{RB-lifetime} ALEPH Collab. D.~Buskulic {\it et ~al.},
                  {\sl Phys. Lett.} {\bf{B313}} (1993) 535; \\
                       OPAL Collab. P.~D.~Acton {\it et ~al.},
                  {\sl Z. Phys.} {\bf{C60}} (1993) 579;
                       D.~Akers {\it et ~al.},
                  {\sl Z. Phys.} {\bf{C65}} (1994) 17;
                         K. Ackerstaff {\it et ~al.},
                     CERN-PPE-96-167, 
                            Submitted to Z. Phys. C. (1996); \\
                       DELPHI Collab. P.~Abreu {\it et ~al.},
                  {\sl Z. Phys.} {\bf{C66}} (1995) 323,
                  {\sl Z. Phys.} {\bf{C70}} (1996) 531.







\bibitem{rbprd} SLD Collab. K.~Abe {\it et ~al.},
                  {\sl Phys. Rev.} {\bf{D53}}, (1996) 1023.


\bibitem{JETSET} T.~Sjostrand, {\sl Comput. Phys. Commun.} {\bf 82}  (1994) 
74. 

\bibitem{CLEO-QQ}  QQ MC code provided by P.~Kim and the CLEO 
                 Collab. Tuning of the $B$ decay modeling is 
               described in SLD Collab., K.~Abe {\it et ~al.}, 
              {\sl SLAC-PUB-7266} (1996), 
               Submitted to {\sl Phys. Rev. Lett.} 



\bibitem{JADE} JADE Collab, S. Bethke {\it et ~al.}, {\sl Phys. Lett.}
 {\bf B213} (1988) 235.

\bibitem{DJNIM} D.~Jackson, 
                  {\sl Nucl. Inst. \& Meth.} {\bf A388} (1997) 247.


\bibitem{ZFITTER} D. Bardin {\it et ~al.}, {\sl CERN-TH 6443/92}, May 1992.


\bibitem{EWWG96} The LEP Electroweak Working Group, 
              "Presentation of the LEP
            Electroweak Heavy Flavour Results for Summer 1996
              Conferences",
              {\sl LEPHF/96-01}, July 1996.

\bibitem{OPALglucc} OPAL Collab. R. Akers {\it et ~al.}, 
                    {\sl Phys. Lett.} {\bf B353} (1995) 595.

\bibitem{Peterson} C. Peterson {\it et ~al.} {\sl Phys. Rev.} {\bf D27} (1983) 105.

\bibitem{Bowler} M. G. Bowler, {\sl Z. Phys.} {\bf C11} (1981) 169.

\bibitem{PDG94} Particle Data Group, {\sl Phys. Rev.} {\bf D50}, Part I (1994).



\bibitem{MK3-DDCY} Mark-III Collab., D. Coffman
{\it et.~al.}, 
                   {\sl Phys  Lett.} {\bf B263} (1991) 135.

\bibitem{HERWIG}  G. Marchesini {\it et ~al.},
                  {\sl Comput. Phys. Commun.} {\bf 67} (1992) 465.

\bibitem{ALEPHNEW} ALEPH Collab. R. Barate {\it et ~al.},
                 {\sl Phys. Lett.}  {\bf B401} (1997) 150, 163. 






}
\end{thebibliography}

\pagebreak

\section*{*List of Authors} 

\bigskip
\begin{center}
\bigskip
%
%
%
  \def\iADEL{$^{(1)}$}
  \def\iBOL{$^{(2)}$}
  \def\iBU{$^{(3)}$}
  \def\iBRUN{$^{(4)}$}
  \def\iUCSB{$^{(5)}$}
  \def\iUCSC{$^{(6)}$}
  \def\iCIN{$^{(7)}$}
  \def\iCSU{$^{(8)}$}
  \def\iCOLO{$^{(9)}$}
  \def\iCOL{$^{(10)}$}
  \def\iFER{$^{(11)}$}
  \def\iFRA{$^{(12)}$}
  \def\iILL{$^{(13)}$}
  \def\iLBL{$^{(14)}$}
  \def\iMIT{$^{(15)}$}
  \def\iMASS{$^{(16)}$}
  \def\iMISS{$^{(17)}$}
  \def\iMOSC{$^{(18)}$}
  \def\iNAG{$^{(19)}$}
  \def\iOREG{$^{(20)}$}
  \def\iPAD{$^{(21)}$}
  \def\iPERU{$^{(22)}$}
  \def\iPISA{$^{(23)}$}
  \def\iRUT{$^{(24)}$}
  \def\iRAL{$^{(25)}$}
  \def\iSOGANG{$^{(26)}$}
  \def\iSOONG{$^{(27)}$}
  \def\iSLAC{$^{(28)}$}
  \def\iTENN{$^{(29)}$}
  \def\iTOH{$^{(30)}$}
  \def\iVAND{$^{(31)}$}
  \def\iWASH{$^{(32)}$}
  \def\iWISC{$^{(33)}$}
  \def\iYALE{$^{(34)}$}
  \def\dead{$^{\dag}$}
  \def\andgen{$^{(a)}$}
  \def\andper{$^{(b)}$}
%
%
\mbox{K. Abe                 \unskip,\iNAG}
\mbox{K. Abe                 \unskip,\iTOH}
\mbox{T. Akagi               \unskip,\iSLAC}
\mbox{N.J. Allen             \unskip,\iBRUN}
\mbox{W.W. Ash               \unskip,\iSLAC$^\dagger$}
\mbox{D. Aston               \unskip,\iSLAC}
\mbox{K.G. Baird             \unskip,\iRUT}
\mbox{C. Baltay              \unskip,\iYALE}
\mbox{H.R. Band              \unskip,\iWISC}
\mbox{M.B. Barakat           \unskip,\iYALE
% ...
}
\it
  \vskip \baselineskip                   
  \vskip \baselineskip                   
%
%
%
  \iADEL
     Adelphi University,
     Garden City, New York 11530 \break
  \iBOL
     INFN Sezione di Bologna,
     I-40126 Bologna, Italy \break
  \iBU
     Boston University,
     Boston, Massachusetts 02215 \break
  \iBRUN
     Brunel University,
     Uxbridge, Middlesex UB8 3PH, United Kingdom \break
  \iUCSB
     University of California at Santa Barbara,
     Santa Barbara, California 93106 \break
  \iUCSC
     University of California at Santa Cruz,
     Santa Cruz, California 95064 \break
  \iCIN
     University of Cincinnati,
     Cincinnati, Ohio 45221 \break
  \iCSU
     Colorado State University,
     Fort Collins, Colorado 80523 \break
  \iCOLO
     University of Colorado,
     Boulder, Colorado 80309 \break
  \iCOL
     Columbia University,
     New York, New York 10027 \break
  \iFER
     INFN Sezione di Ferrara and Universit\`a di Ferrara,
     I-44100 Ferrara, Italy \break
  \iFRA
     INFN  Lab. Nazionali di Frascati,
     I-00044 Frascati, Italy \break
  \iILL
     University of Illinois,
     Urbana, Illinois 61801 \break
  \iLBL
     E.O. Lawrence Berkeley Laboratory, University of California,
     Berkeley, California 94720 \break
  \iMIT
     Massachusetts Institute of Technology,
     Cambridge, Massachusetts 02139 \break
  \iMASS
     University of Massachusetts,
     Amherst, Massachusetts 01003 \break
  \iMISS
     University of Mississippi,
     University, Mississippi  38677 \break
  \iMOSC
    Moscow State University,
    Institute of Nuclear Physics
    119899 Moscow, Russia    \break
  \iNAG
     Nagoya University,
     Chikusa-ku, Nagoya 464 Japan  \break
  \iOREG
     University of Oregon,
     Eugene, Oregon 97403 \break
  \iPAD
     INFN Sezione di Padova and Universit\`a di Padova,
     I-35100 Padova, Italy \break
  \iPERU
     INFN Sezione di Perugia and Universit\`a di Perugia,
     I-06100 Perugia, Italy \break
  \iPISA
     INFN Sezione di Pisa and Universit\`a di Pisa,
     I-56100 Pisa, Italy \break
  \iRUT
     Rutgers University,
     Piscataway, New Jersey 08855 \break
  \iRAL
     Rutherford Appleton Laboratory,
     Chilton, Didcot, Oxon OX11 0QX United Kingdom \break
  \iSOGANG
     Sogang University,
     Seoul, Korea \break
  \iSOONG
     Soongsil University,
     Seoul, Korea  156-743 \break
  \iSLAC
     Stanford Linear Accelerator Center, Stanford University,
     Stanford, California 94309 \break
  \iTENN
     University of Tennessee,
     Knoxville, Tennessee 37996 \break
  \iTOH
     Tohoku University,
     Sendai 980 Japan \break
  \iVAND
     Vanderbilt University,
     Nashville, Tennessee 37235 \break
  \iWASH
     University of Washington,
     Seattle, Washington 98195 \break
  \iWISC
     University of Wisconsin,
     Madison, Wisconsin 53706 \break
  \iYALE
     Yale University,
     New Haven, Connecticut 06511 \break
  \dead
     Deceased \break
  \andgen
     Also at the Universit\`a di Genova \break
  \andper
     Also at the Universit\`a di Perugia \break
\rm
%

\end{center}


\begin{table}[htb]
\begin{center}
\begin{tabular}{| l | r | }
\hline
{\bf Light Quark Systematic ($\epsilon_{uds}$)} & $\delta R_b$  \\
\hline
\glubb\ 0.31$\pm$0.11\%                & -0.00033  \\
\glucc\ 2.38$\pm$0.48\%                & -0.00004  \\
$K^0$ production $\pm$10\%             & -0.00003  \\
$\Lambda$ production $\pm$10\%         & -0.00002  \\
\hline 
Total $uds$ physics systematic         & {\bf 0.00034}  \\
\hline
\hline 
{\bf Charm Systematic ($\epsilon_c$)}  & $\delta R_b$\\
\hline
$D^+$ production  $0.259\pm0.028$      & -0.00011 \\
$D_s$ production  $0.115\pm0.037$      & -0.00005 \\
$c$-baryon production $0.074 \pm0.029$  &  0.00011 \\
c-frag. $\langle x_E \rangle_D=0.482 \pm0.008$  & -0.00006 \\
c-frag. function shape                    & -0.00001 \\
\hline 
$D^0$ lifetime $0.415\pm0.004$ ps                    & -0.00003 \\
$D^+$ lifetime $1.057\pm0.015$ ps                    & -0.00001 \\
$D_s$ lifetime $0.467\pm0.017$ ps                    & -0.00002 \\
$\Lambda_c$ lifetime $0.200\pm0.011$ ps              & -0.00001 \\
\hline 
$D^0$ decay $\langle N_{ch} \rangle =2.54\pm0.05$& -0.00006 \\
$D^+$ decay $\langle N_{ch} \rangle =2.50\pm0.06$& -0.00006 \\
$D_s$ decay $\langle N_{ch} \rangle =2.65\pm0.33$& -0.00009 \\
\hline                                                            
$D^0\ra K^0$ production $ 0.401 \pm0.059$     & +0.00015 \\
$D^+\ra K^0$ production $ 0.646 \pm0.078$     & +0.00020 \\
$D_s\ra K^0$ production $ 0.380 \pm0.06$       & +0.00002 \\
\hline                                                             
$D^0$ decay no-$\pi^0$ frac. $0.370\pm0.037$       & +0.00005 \\
$D^+$ decay no-$\pi^0$ frac. $0.499\pm0.050$       & -0.00008 \\
$D_s$ decay no-$\pi^0$ frac. $0.352\pm0.035$       &$<$0.00001\\
\hline 
Total Charm Physics systematic                        &  {\bf 0.00033} \\
\hline
\hline 
{\bf B decay modeling ($\lamb$)}                     & $\delta R_b$\\
\hline
$B$ lifetime $\pm0.05$~ps      & 0.00004\\
$B$ decay   $\langle N_{ch} \rangle = 5.73\pm0.35$  & 0.00003 \\
$b$ fragmentation               & 0.00019 \\
$\Lambda_b$ production fraciton $ 0.074 \pm0.03$           & 0.00008 \\
Hard gluon radiation                     & 0.00008 \\
$B$ momentum correlation            & 0.00029 \\
$b$-tag $\cos \theta$ dependency                  & 0.00001 \\
\hline 
Total \bbbar\ Physics systematic    & {\bf 0.00038}  \\
\hline
\hline 
{\bf Detector Systematic}       &$\delta R_b$\\
\hline
Tracking resolution               & 0.00096\\
Tracking efficiency               & 0.00040 \\
$\langle IP \rangle_{xy}$ tail   & 0.00010 \\
MC statistics                     & 0.00091 \\
Event selection bias              & 0.00028 \\
\hline 
Total detector and MC             & {\bf 0.00141}  \\
\hline \hline
{\bf $R_c=0.171 \pm 0.006$ }          & {\bf 0.00021}   \\
\hline \hline 
{\bf Total (excl. $R_c$)} & {\bf 0.00154} \\
\hline \hline 
\end{tabular}
\caption{ 
Summary of systematic uncertainties for the $ \Mtag > 2.0$~GeV/$c^2$ cut.}
\label{tab:syst}
\end{center}
\end{table}

\end{document}